\documentclass[
  journal=pasa,
  manuscript=research-paper, %% or "review"
  year=2020,
  volume=37,
]{cup-journal}

\usepackage{microtype,siunitx,booktabs}
\usepackage{amsmath,amssymb}
\usepackage{hyperref}
\hypersetup{colorlinks,citecolor=blue,linkcolor=blue,urlcolor=blue}
\usepackage{aas-macros}
\usepackage{threeparttablex}
\usepackage{longtable}

\title{The transience and persistence of high optical polarization state in beamed radio quasars}
\author{Krishan Chand}
\affiliation{Aryabhatta Research Institute of Observational Sciences (ARIES), Manora Peak, Nainital 263002, India}
\alsoaffiliation{Department of Physics, Kumaun University, Nainital 263002, India}
\email[Krishan Chand]{krishanchand007.kc@gmail.com}

\author{Gopal-Krishna}
\affiliation{UM-DAE Centre for Excellence in Basic Sciences, Vidyanagari, Mumbai 400098, India}
\author{Amitesh Omar}
\affiliation{Aryabhatta Research Institute of Observational Sciences (ARIES), Manora Peak, Nainital 263002, India}

\author{Hum Chand}
\affiliation{Department of Physics and Astronomical Sciences, Central University of Himachal Pradesh (CUHP), Dharamshala 176215, India}

\author{P. S. Bisht}
\affiliation{Department of Physics, Soban Singh Jeena University, Almora 263601, India}

\doi{10.1017/pasa.2020.32}

\received {dd Mmm YYYY}
\revised  {dd Mmm YYYY}
\accepted {dd Mmm YYYY}
\published{22 September 2020}

\begin{document}

\begin{abstract}
We examine the long-term stability (on decade-like time scales) of {optical `high polarization' (HP)} state {with} ${p_{opt}}$ ${> 3\%}$, which commonly occurs in flat-spectrum (i.e., beamed) radio quasars (FSRQs)
%, with a high fractional optical polarization ($p_{opt}$ > 3\%) being 
{and is} a prominent marker of blazar state. Using this clue, roughly a quarter of the FSRQ population has been reported to undergo {HP} $\leftrightarrow$ {non-HP} state transition on year-like time scales. This work examines the extent to which {HP (i.e., blazar)} state can endure in a FSRQ, despite these `frequent' state transitions. This is the first attempt to verify, using purely opto-polarimetric data for a much enlarged sample of blazars, the recent curious finding that blazar state in individual quasars persists for {\it at least} a few decades, despite its changing/swinging observed fairly commonly on year-like time scales. The present analysis is based on a well-defined sample of {83} radio quasars, extracted from the opto-polarimetric survey RoboPol (2013-17), for which old opto-polarimetric data taken prior to 1990 could be found in the literature. By a source-wise comparison of these two datasets of the same observable ($p_{opt}$), we find that $\sim$ {90}\% of the {63} quasars found in blazar state in our RoboPol sample, were also observed to be in that state about 3 decades before. On the other hand, within the RoboPol survey itself, we find that roughly a quarter of the {blazars} in our sample migrated to the other polarization state on year-like time scales, by crossing the customary $p_{opt}$ = 3\% threshold. Evidently, these relatively frequent transitions (in either direction) do not curtail the propensity of a radio quasar to retain its blazar {(i.e., HP)} state for at least a few decades. The observed transitions/swings of polarization state are probably {manifestation of} transient processes, like ejections of synchrotron plasma blobs (VLBI radio knots) from the active nucleus.
\end{abstract}
 
\section{Introduction}
\label{Introduction}
A high and variable fractional optical linear polarization ($p_{opt}$) is a well-known characteristic of blazars \citep[e.g.,][]{Strittmatter1972ApJ...175L...7S}, which is a subset of flat-spectrum radio quasars (FSRQs) whose flux density at not only centimetric but also at optical and possibly at even shorter wavelengths is dominated by a synchrotron jet relativistically beamed in our direction \citep[e.g.,][]{Blandford1978bllo.conf..328B,Begelman1984RvMP...56..255B,Antonucci1993ARA&A..31..473A,Urry1995PASP..107..803U}. 
Guided by the limiting polarization of up to $\sim$ 2\% that can arise from scattering (with extremely few outlier QSOs), two polarization classes have been defined for quasars, separated at $p_{opt}$ = 3\% \citep[e.g.,][]{Stockman1984ApJ...279..485S}.
%,Antonucci1985ApJ...297..621A}.
These are: (i) high-polarization quasars (HPQ, $p_{opt}$ $>$ 3\%) which are nearly always radio-loud {and are referred to here as blazars, dominated by relativistically beamed radio-optical synchrotron radiation 
(e.g., \citealp{Wills1992ApJ...398..454W}),} 
and (ii) low-polarization radio quasars (LPRQ).
%The blazar subset among FSRQs is identified with the HPQ class, and it also includes BL Lac objects whose optical/UV spectra exhibit at most very weak emission lines (e.g.,  the review by \citealp{Urry1995PASP..107..803U}). 

{The opto-polarimetric division} begs the question about trans-migration of individual FSRQs between the two polarization classes. Early attempts to examine this were reported by \citet{Fungmann1988AA...205...86F}, \citet{Kuehr1990AJ.....99....1K} and \citet{Impey1990ApJ...354..124I}, hereafter, IT90. Using optical polarimetric observations separated by month-like time scales, \citet{Fungmann1988AA...205...86F} inferred that, statistically, a FSRQ exists in a blazar state with a duty cycle of $\sim$ 2/3, so that at any time, nearly two-thirds of a sample of FSRQs exhibits blazar characteristics {(HP)}, as also found by IT90. This poses the {next} question {which is} about the typical time-scale on which the above duty cycle operates. An early quantitative check was reported in IT90, based on an assortment of multi-epoch (mostly two-epoch) optical polarimetry of 41 FSRQs. It showed that nearly a quarter of the FSRQs jumped the $p_{opt}$ = 3\% threshold (in either direction), on year-like time scales. This would indicate that over a much longer, say decade-like, time interval the memberships of the HPQ and LPRQ subsets in a sample of FSRQs would undergo a drastic churn, even though their relative numerical strengths remain unchanged. 
One study bearing on this question employed strong intranight optical variability (INOV), a well-established signature of blazar activity (\citealp{Goyal2012AA...544A..37G}, \citealp{GoyalA2013MNRAS.435.1300G}, also, \citealp{Gopal-Krishna2018BSRSL..87..281G}). \citet{Goyal2012AA...544A..37G} specifically investigated the dependence of INOV of FSRQs on optical polarization, by carrying out sensitive and densely sampled intranight optical monitoring of 9 HPQs and 12 LPRQs. Remarkably, the HPQ subset showed strong INOV (i.e., amplitude $\psi$ $>$ 4\%) on 11 out of 29 nights, in stark contrast to the LPRQs for which strong INOV was observed on just 1 out of 44 nights. {This demonstrates that the key blazar characteristic of rapid optical variability is primarily a property of HPQs. This is also consistent with the recent study of 8 FSRQs, based on a 10-year long monitoring at Steward observatory of their optical flux and polarization (\citealp{Pandey2022MNRAS.510.1809P}). This study revealed that in an overwhelming majority of cases ($\sim$ 90\%), changes in optical flux 
%(which is largely a measure of relativistic beaming) 
on day-to-month-like time scales are positively correlated with the degree of optical polarization, again in accord with the notion that a high optical polarization is associated with relativistically beamed synchrotron component, a key marker of blazar activity (e.g., \citealp{Wills1992ApJ...398..454W}; \citealp{Marscher2016Galax...4...37M}). This correlation is widely understood in terms of the, so called, `shock-in-jet' model of relativistic blazar jets, wherein both particle acceleration and magnetic field alignment take place at shocks within the jet flow (\citealp{Blandford1979ApJ...232...34B}; \citealp{Hughes1985ApJ...298..301H}; \citealp{Marscher1985ApJ...298..114M}).}

Another striking aspect highlighted in {the afore-mentioned study by \citet{Goyal2012AA...544A..37G} is} that the grossly divergent INOV behaviour of the two FSRQ subsets was observed even though their HPQ/LPRQ classification had been made more than 2 decades prior to the INOV campaign, indicating a long-term memory of the polarization class. 
Recently, we revisited this curious finding about the time-scale over which the HPQ (i.e., blazar) state of a FSRQ persists (\citealp{Krishan2022MNRAS.516L..18C}, Paper I), by checking for the {HP} state among individual members of a well-defined sample of 80 FSRQs, selected from IT90, out of which 49 had been confirmed to be in {HP
(blazar)} state during 1970-80s, through opto-polarimetry. For each FSRQ, we then checked for blazar state around 2020, based on flux variability in its high-quality optical light-curves obtained under the ongoing Zwicky Transient Facility (ZTF)\footnote{The database of the ongoing ZTF project, \url{https://www.ztf.caltech.edu}} project which was launched in 2018 and presents an unprecedented combination of sensitivity and cadence \citep{Belim2019PASP..131a8002B}.
It was thus found in Paper I that over the time baseline of $\sim$ 4 decades, not more than $\sim 10\%$ of the FSRQs showed a clear sign of transition to {(or from) blazar state}. Paper I also reported supporting evidence for this rather unexpected result, utilising the `recent' polarization state classification based on the opto-polarimetric measurements (instead of optical flux variability) {that are} available for 32 out of the {total} sample of 80 FSRQs, in the archives of the RoboPol survey conducted during 2013-17 (\citealp{Blinov2021MNRAS.501.3715B}, hereafter, B21). In this paper we employ a 2.6 times larger quasar sample which has enabled a statistically much more robust test of the persistence of blazar{/HP} state among quasars, while ensuring that the blazar state at {\it both} ends of the time baseline is {assigned in terms of} opto-polarimetric data alone.

\section{The opto-polarimetric sample of quasars}
\label{The quasar sample}

Out of the total 222 radio quasars in the RoboPol survey (B21), 210 have been marked as blazar/FSRQ in the ROMA-BZCAT catalogue \citep{Massaro2015ApSS.357...75M}. We 
were able to assign to 83 of them `old' opto-polarimetric class through an extensive (though not complete) search of the literature published before 1995 (data {acquired} no later than 1990), albeit from sensitivity considerations data taken before 1968 were not considered (see IT90). These 83 quasars constitute our RoboPol sample 
(Table \ref{tab1}). The AGN classes mentioned in col. 2, taken from the ROMA-BZCAT catalogue are based on multiple criteria{, namely}: (i) radio detection and radio core-dominance/spectral-index, (ii) optical spectrum, and (iii) X-ray luminosity. For {these reasons}, this classification does not bear a one-to-one correspondence with the blazar classification scheme adopted in the present work, 
%where the criterion adopted for classifying the radio quasars is more sharply defined, 
which is based solely on the afore-mentioned $p_{opt} > 3\%$ criterion. 
%Therefore, in order to avoid confusion, we shall refer to the various classes mentioned in col. 2 of Table \ref{tab1} by a single term ``Blazar Candidate (BZC)'' and then employ the $p_{opt}$ > 3\% criterion for confirming the blazar nature.

Using the published opto-polarimetric measurements made a few decades ago (Table \ref{tab1}, col. 6), and then again in recent years (2013-17, RoboPol) (see, Table \ref{tab1}, col. 9), we have assigned to each quasar polarization categories `old' and `new', each based on the maximum observed polarization, $p_{opt}(max)$. The maximum value of $p_{opt}$ is preferred over averaged value, as this reduces the chance of missing out genuine blazars/HPQs since their polarization is known to vary and hence may sometimes dip below the defining threshold of $3\%$ (section \ref{Introduction}), particularly because for a majority of the quasars, at most a couple of polarization measurements are available in the old literature. In relatively rare cases where the old literature lists multiple values of $p_{opt}$ for a quasar, all consistent with the same classification (HPQ, or LPRQ), we have selected one representative value of {$p_{opt}$} out of them, by applying a judicious mix of criteria. These criteria are: the availability (and smallness) of the rms error and of the epoch of the measurement, giving a higher weightage to older measurements with a view to extend the time baseline (see the entries in bold-face, in col. 6 in Table \ref{tab1}). 
It may be noted that for making a proper comparison of the RoboPol measurements with those found in the old literature where the available number of polarization measurements per quasar seldom exceeds 2, we have also limited the number of RoboPol measurements accordingly. Thus, for each of the 58 of our 83 quasars, for which more than 3 RoboPol measurements are available (N > 3), we have selected 3 out of the available measurements in an unbiased manner, using a random-number generator (col. 9 in Table \ref{tab1}). As for the old data, the RoboPol-based polarimetric classification of each quasar in our sample was then carried out based on the highest 
$p_{opt}$ ($p_{new}(max)$ listed in Table \ref{tab1}), yielding {59} HPQs, {18} LPRQs and 6 border-line cases {(4 probable HPQs and 2 probable LPRQs)} for which $p_{new} (max)$ is consistent with either classification within the quoted rms error (Table \ref{tab1}). Note that for {8} of the LPRQs, the classification is based on a single available measurement of $p_{new}$. Thus, for several of these (6 + 8 =) 14 quasars, the `new' classification may actually fall under the HPQ category. %We also note that for many sources in our sample, with a large number of RoboPol polarimetric measurements (say, N > 10), the published {\it average} value of $p_{opt}$ may well be a more reliable indicator of the polarization class than $p_{new} (max)$ used here. The 7 quasars for which these two estimates of the (RoboPol) polarization class are found to differ have been superscripted with a ``$\dagger$" in col. 11 of Table \ref{tab1}.

 %\floatsetup[table]{capposition=top}
 \begin{table*}
  \centering
\begin{minipage}{170mm}
    \caption{The basic optical data and the polarization classifications for the present sample of {83} radio quasars (section \ref{The quasar sample}). Only the first 2 sources are shown here, the full table is available in the on-line material.}
    \label{tab1}
    \resizebox{\textwidth}{!}{%
    
\begin{tabular}{ccccccccccc}
  %\scriptsize
  \hline\\
\multicolumn{1}{c}{Source}&class &\multicolumn{1}{c}{{\it z}}  &\multicolumn{1}{c}{App. mag.} &\multicolumn{1}{c}{App. mag.}&\multicolumn{1}{c}{$p_{opt}$}&Pol. &Epoch  &$p_{opt}$ (\%)&N&Pol.\\
   SDSS name               &     &                        & (SIMBAD)                     &(SDSS)                       &(\%)                          & class&                           &   RoboPol&&class\\
                           &     &                       & {\textcolor {blue}B}         &                             &                 &             &                            &$p1$ (JD*)      &&(RoboPol)\\
                           &     &                        &{\textcolor {cyan}V}          &{\textcolor {Green}g}       &                  &            &                             &$p2$ (JD*)      &&\\
   &     &                        &{\textcolor {red}R}           &{\textcolor {red}r}         &                              &           &                  &$p3$ (JD*)      &&\\
                           &                          &                           &                         &                          &  &&                     &mean $p$       & & \\
(1)                        &   (2)                       &(3)                           &(4)                         & (5)                          &(6)  &(7)                     &(8)       & (9)& (10)&(11)\\\\
 \hline\\
0003-066		&BZB  	& 0.347	& {\textcolor {blue}{19.10}}	& 	& ${1.4\pm1.2}$ {$^{F88}$} &(HPQ)&1984	 	&$34.4\pm0.9$ (6599)&15	&HPQ\\
J000613.89-062335.33	& 	& 0.347 	& {\textcolor {cyan}{---}}	& {\textcolor {Green}{18.56}}	&${\mathbf{3.5\pm1.6}}$ {$^{F88}$}&	& 1984	&$26.1\pm1.7$ (6878)&	&\\
& 	& 	 	&{\textcolor {red}{17.14}}	& {\textcolor {red}{17.87}}& ${3.5\pm1.0}$ {$^{K90}$}&	&1987 	&$25.3\pm0.8$ (6872)	&\\
& 	& 		&{\textcolor {red}{}}	& {\textcolor {red}{}}& &	&  	&$<21.20>$	&\\\\
%& 	& 	& 	&{\textcolor {red}{}}	& {\textcolor {red}{}}&	&  	&	&\\
%& 	& 	& 	&{\textcolor {red}{}}	& {\textcolor {red}{}}& 	&  	&	&\\
%			&& 	& 	& 	{\textcolor {red}{}}	& {\textcolor {red}{}}& 	&  	&	&\\\\

 0014+813		&BZQ  	& 3.366	& {\textcolor {blue}{17.60}}	& 	& ${1.1\pm0.1}${$^{K83}$}&LPRQ &1982	 	&$1.8\pm0.5$ (6942)&30	&LPRQ\\
J001708.48+813508.14	& 	& 3.366 	& {\textcolor {cyan}{16.52}}	& {\textcolor {Green}{---}}&	&	& 	&$1.0\pm0.4$ (7333)&	&\\
& 	& 	 	&{\textcolor {red}{15.95}}	& {\textcolor {red}{---}}& &	& 	&$0.5\pm0.5$ (7330)	&\\
& 	& 	 	&{\textcolor {red}{}}	& {\textcolor {red}{}}& 	& &	&$<0.80>$ 	&\\\\

%0133+476             & BZQ          &0.859          &{\textcolor {blue}{---}}    &	            & ${\mathbf{20.8\pm0.7}}${$^{I90}$}    &1986   &$14.6\pm0.8$ (6591)&64&HPQ\\     
%J013658.59+475129.10 &             & 0.859  &       {\textcolor {cyan}{18.00}}    &{\textcolor {Green}{18.70}}           &        &      &$6.4\pm1.5$ (7617)\\
%&             &         &       {\textcolor {red}{19.25}}    &{\textcolor {red}{18.13}}                   & &     &$4.1\pm1.5$ (7607)\\
%&             &         &                                    &                                            & &     &$<8.75>$\\\\

---            &---       & ---  &---      &---   &---&---    &---           &---&---        &---		\\
%---            &---       & ---        &---   &---&---    &---           &---&---        &---	&---	&---   &---  & ---     &\\

\hline\\
\multicolumn{11}{l}{{\bf Col. (2):} The source class is taken from the 5th edition of ROMA-BZCAT \citep{Massaro2015ApSS.357...75M} where the BZB stands for BL Lac type of blazar,}\\
\multicolumn{11}{l}{ BZQ stands for FSRQ type blazar, BZG: BL Lac-galaxy dominated and BZU: blazars of Uncertain type; {\bf Col. (3):} Redshift, the upper value is from} \\
%\multicolumn{12}{l}{Col. (3): Radio spectral index ($f_{\nu} \propto \nu^{\alpha}$), taken from \citet{Kuehr1981AAS...45..367K};}\\
\multicolumn{11}{l}{  NED and the lower value is from \citet{Blinov2021MNRAS.501.3715B}. The redshift marked with `a' and `b' is taken from \citet{Tarno2020ApJS..250....1T} and }\\
\multicolumn{11}{l}{\citet{Landoni2020ApJS..250...37L} respectively; {\bf Col. (6):} The reference codes for polarization are shown as the superscripts, where A80 stands for }\\
\multicolumn{11}{l}{\citet{Angel1980ARAA..18..321A}, A84: \citet{Antonucci1984ApJ...278..499A}, B90: \citet{Ballard1990MNRAS.243..640B}, Bi81: \citet{Biermann1981ApJ...247L..53B}, B86: \citet{Brindle1986MNRAS.221..739B}, C88:}\\
\multicolumn{11}{l}{\citet{Courvoisier1988Natur.335..330C}, F88: \citet{Fugmann1988AAS...76..145F}, I82: \citet{Impey1982MNRAS.198....1I}, I88: \citet{Impey1988ApJ...333..666I}, I90: \citet{Impey1990ApJ...354..124I}, }\\
\multicolumn{11}{l}{I91: \citet{Impey1991ApJ...375...46I}, J93: \citet{Jannuzi1993ApJS...85..265J}, K76: \citet{Kinman1976ApJ...205....1K}, K83: \citet{Kuhr1983ApJ...275L..33K}, K90: \citet{Kuehr1990AJ.....99....1K}, M83:  }\\
\multicolumn{11}{l}{\citet{Martin1983ApJ...266..470M}, M90: \citet {Mead1990AAS...83..183M}, M75: \citet{Miller1975ApJ...200L..55M}, M81: \citet{Moore1981ApJ...243...60M}, M84: \citet{Moore1984ApJ...279..465M}, P83: }\\
\multicolumn{11}{l}{\citet{Puschell1983ApJ...265..625P}, Si85: \citet{Sitko1985ApJS...59..323S}, S86: \citet{Smith1986ApJ...305..484S}, S87: \citet{Smith1987ApJS...64..459S}, S88: \citet{Smith1988ApJ...326L..39S}, S78: }\\ 
\multicolumn{11}{l}{\citet{Stockman1978ApJ...220L..67S}, S84: \citet{Stockman1984ApJ...279..485S}, W80: \citet{Wills1980AJ.....85.1555W} and W92: \citet{Wills1992ApJ...398..454W}; {\bf Col. (7):} polarization class of the}\\
% and H98 for \citet{Hutsemekers1998AA...332..410H};\\
\multicolumn{11}{l}{ source ($p_{opt}$(max) $\leq$ 3\% for LPRQs and $>$ 3\% for HPQs). The polarization class in parentheses is for border-line cases for which the quoted one-$\sigma$ }\\
\multicolumn{11}{l}{error on $p_{max}$ would push the source to the other polarization class; {\bf Col. (8):} Epoch of the $p_{opt}$ measurement; {\bf Col. (9):} RoboPol measured $p_{opt}$, }\\
\multicolumn{11}{l}{taken from \citet{Blinov2021MNRAS.501.3715B}. The JD in parentheses marked with `*' corresponds to Julian date minus 2450000. The parameters $p1$, $p2$ and $p3$ are  }\\
\multicolumn{11}{l}{explained in section \ref{The quasar sample}; {\bf Col. (10):} Number of RoboPol polarization measurements; {\bf Col. (11):} polarization class of the source ($p_{opt}$(max) $\leq$ 3\% for LPRQs and}\\
\multicolumn{11}{l}{$>$ 3\% for HPQs). The polarization class in parentheses is for those six sources for which the quoted one-$\sigma$ error on $p_{max}$ would push the source to the}\\
\multicolumn{11}{l}{ other polarization class. Note that all these six sources have the number of RoboPol measurements N < 5.}\\% The sources for which the polarization classes }\\
%\multicolumn{11}{l}{based on $p_{opt}$(max) and average $p$ are different, have been marked with `$\dagger$' superscipted on the polarization class mentioned in col. 11.}\\
%\multicolumn{11}{l}{}\\
% \insertTableNotes\\

\end{tabular}
 } 
 \end{minipage}
\end{table*}

\section{Discussion}
\label{Discussion}

%{\BF I DO NOT FIND CAPTION TO FIG. 1. THE CAPTION SHOUOLD MENTION WHICH SAMPLE IS PLOTTED IN THE DIAGRAM}

\begin{figure}

  \includegraphics[width=9.5cm,height=9.5cm,trim=1.2cm 0cm 0.0cm 2.5cm,clip]{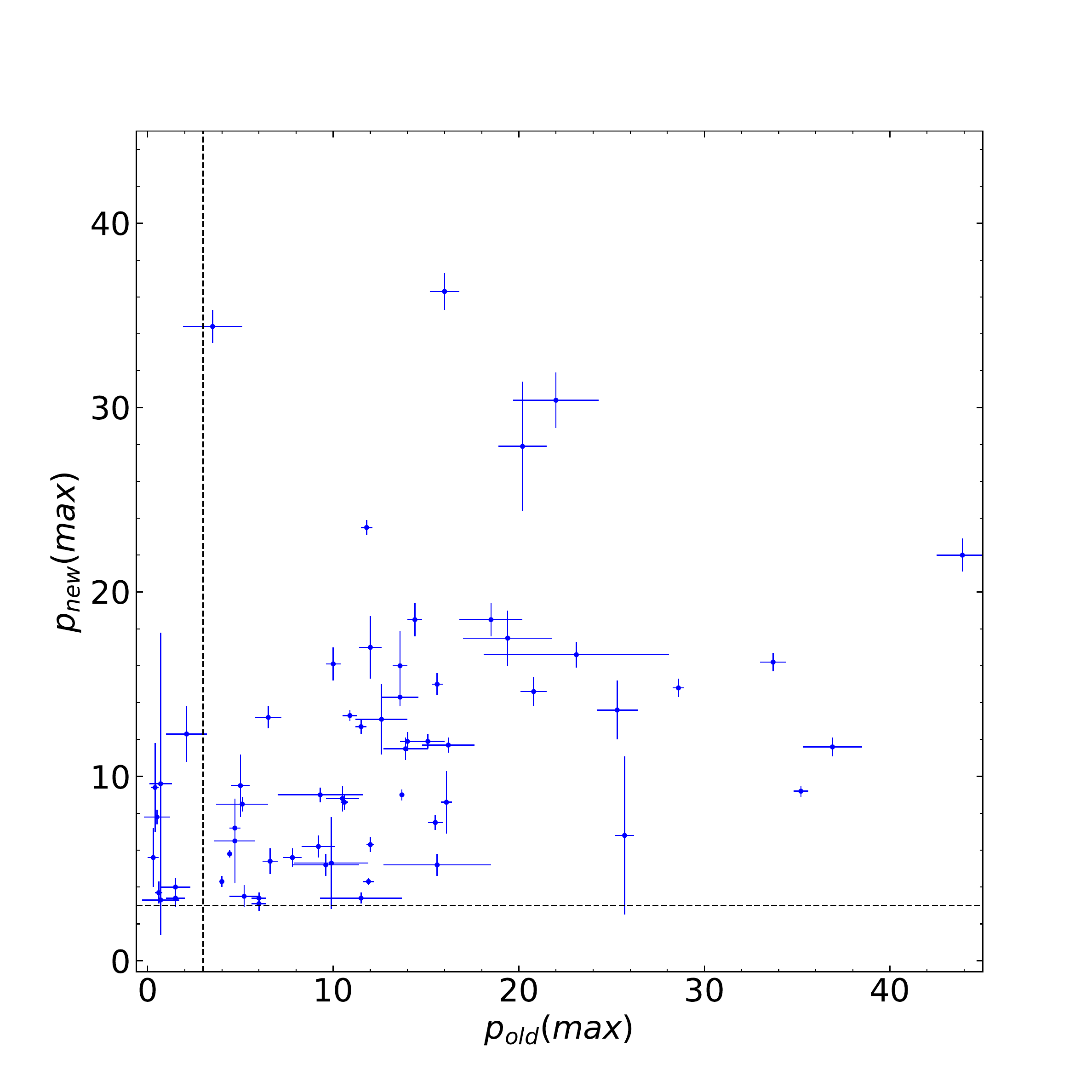}\\
  \vspace{-0.25in}
  \caption{Plot of maximum reported polarization from the old literature ($p_{old}(max)$, shown in bold-face in col. 6 of Table \ref{tab1}) and from RoboPol survey ($p_{new}(max)$, listed in col. 9 of Table \ref{tab1}), for the 63 HPQs, including the 4 border-line HPQs (see Table \ref{tab1}).}
  \label{fig:2}
\end{figure}

Fig. \ref{fig:2} shows a plot of $p_{opt}(max)$ at the two ends of the time baseline, for the {63} HPQs in our RoboPol sample, including the 4 border-line HPQs ({1012+232, 1514-241, {\underline{1638+398}} and 2150+173}) {(see section \ref{The quasar sample}, {also} Table \ref{tab1})}. Thus, we plot $p_{old}(max)$ (from the 1973-89 literature, shown in bold-face in col. 6 of Table \ref{tab1}) against $p_{new}(max)$ (from RoboPol survey 2013-17, col. 9). 
%{\bf (Note: There are 6 borderline sources. 4 out of 6 are designated as (HPQs) and 2 are (LPRQs). In the number 63 the former 4 have been taken which are (HPQs)+59)}. 
%It is evident that only between 2 to {\bf6} HPQs identified in the RoboPol survey were possibly in LPRQ mode a few decades before. 
Based on this diagram, we discuss below individually the candidates for polarization state transition over this time interval of $\sim 3.5$ decades. {In tandem, we shall also examine the role of the thermal UV emission arising from the accretion disk, which would increasingly contribute to the observed optical flux of high-$z$ quasars
(see, e.g., \citealp{Wills1992ApJ...398..454W}), expectedly moderating their observed optical flux variability as well as fractional polarization. For the purpose of this check, we shall thus limit our analysis to the $z < 1$ members of our HPQ sample, numbering 54 including the 3 border-line HPQs (for identifying $z > 1$ HPQs, their names appear underlined in this section).}

\subsection{Cases of long-term change in the polarization state}

\subsubsection{The RoboPol HPQs, i.e., blazars}
\label{The RoboPol HPQs}
From the $p_{old}(max)$ vs $p_{new}(max)$ plot for the {63} HPQs in our RoboPol sample (Fig. \ref{fig:2}; Table \ref{tab1}), including the 4 border-line cases mentioned above, and using the information provided in Table \ref{tab1}, we have shortlisted the following {9} HPQs as candidates for polarization state transition.

0748+126: This HPQ with two measurements of $p_{new}$ (= ${12.3\pm1.5\%}$ and ${4.2\pm1.1\%}$) had been found to be a confirmed LPRQ, based on all 3 available measurements during 1984-86 (Table \ref{tab1}). Hence, {it is} a clear case of polarization state transition.

0827+243: With $p_{new}(max)$ = $4.0\pm0.5\%$, this source qualifies as an HPQ, and also as a case for polarization state transition, having been earlier recorded as an LPRQ on the basis of two available measurements during 1979-80. Note, however, that this quasar is only a `probable' case {of polarization state transition} since its average $p_{new}$, based on 13 RoboPol measurements is clearly below $3\%$.

{\underline{0850+581}}: With $p_{new}(max)$ = $9.4\pm2.4\%$, this is a confirmed RoboPol HPQ. However, it is only a `probable' case of polarization state transition, since its LPRQ classification in the old literature is based on a single measurement, due to which the possibility of its then being an HPQ cannot be ignored.

1012+232: With $p_{new}(max)$ = $9.6\pm8.2\%$, its HPQ classification is highly uncertain and even its LPRQ classification in the old literature (1985) is based on single measurement, $p_{old}$ = $0.7\pm0.6$ (Table \ref{tab1}). Therefore, {it is} at most a `probable' case of state transition.

{\underline{1551+130}}: With $p_{new}(max)$ = $3.3\pm0.7\%$, this border-line HPQ, is likely to be actually an LPRQ, given its average $p_{opt}$ of 1.83\%, based on 47 RoboPol measurements. {In the old literature (1992), this quasar is classified as an LPRQ ($p_{old}$ = $0.7\pm1.0\%$), making it} an unlikely case of polarization state transition.%Also, its LPRQ classification in the old literature (1992) is based on single measurement ($p_{old}$ = $0.7\pm1.0\%$). Hence, an unlikely case of polarization state transition.

{\underline{1633+382}}: The polarimetric state transition of this HPQ with $p_{new}$ consistently remaining above 3\%, is confirmed, since all 5 measurements available in the old literature gave $p_{old}$ $< 3\%$ (i.e., LPRQ, see Table \ref{tab1}). Its transition has also been highlighted in \citet{Lister2000ApJ...541...66L}. 

{\underline{1954+513}}: With $p_{new}(max)$ = $3.4\pm0.5\%$, this HPQ, is likely to be actually an LPRQ, given its average $p_{opt}$ of 2.78\%, based on 17 RoboPol measurements. Also, its LPRQ classification in the old literature (1980) is based on single measurement ($p_{old}$ = $1.5\pm0.5\%$). Hence, an unlikely case of polarization state transition.

2021+614: With two measurements of $p_{new}$ (= $5.6\pm1.6\%$ and $4.7\pm1.3\%$), the source is a confirmed RoboPol HPQ. However, its LPRQ classification in the old literature (1986) is based on single measurement ($p_{old}$ = $0.3\pm0.3\%$). Hence, it can only be termed as a `probable' case of polarization state transition.

2145+067: This quasar, with $p_{new} (max) = 3.7\pm0.6\%$ is a border-line HPQ, particularly since its available 70 RoboPol measurements give an average $p_{opt}$ = 1.28\%. Thus, even though it is a confirmed LPRQ in the old literature, it remains at best a weak case of polarization state transition.

In summary, we find that out of the {63} HPQs in our RoboPol sample, just 2 are confirmed cases and 4 are `probable' 
%and {3} are `weak/unlikely' 
cases of polarization state transition over the time interval of $\sim 3$ decades preceding the RoboPol survey. {Thus,} the frequency of polarimetric state transition is $\sim$ {3\%}, and within 10\% even if the 4 `probable' cases are accepted. This is consistent with the upper limit of $\sim 10\%$ deduced in Paper I, based albeit on a 2.3 times smaller RoboPol sample of 27 HPQs. {A consistent result is found when we consider just the $z < 1$ subset of the 63 HPQs. It is found that out of the total 54 such HPQs, one (0748+126) is a confirmed case of transition and another three (0827+243, 1012+232 and 2021+614) are probable cases. Thus, taken together, at most four (i.e., $\sim$ 7\%) of the 54 HPQs with $z < 1$ have undergone polarimetric state transition over the time interval of $\sim 3$ decades preceding the RoboPol survey.}

%({\bf Note: Based on $p_{new}$(max) in Paper I, out of 32 sources, 27 are HPQs and 5 are LPRQs.})

\subsubsection{The RoboPol LPRQs}
\label{LPRQs}
Bearing in mind the fluctuating polarization of HPQs, and broadly in conformity with the analysis of IT90, we shall accept a source as a confirmed LPRQ only provided its $p_{opt}$ is consistently found to be $\leq$ 3\% in all available (at least two) measurements. Considering $p_{new}$ values given in col. 9 of Table \ref{tab1}, this condition is satisfied by 11 quasars having N $\geq$ 2. These are:
 {\underline{0014+813}}, {\underline{0333+321}}, 0403-132, {\underline{0833+585}}, {\underline{0836+710}}, 1426+428, {\underline{1606+105}}, 1637+574, 1652+398, 1928+738 and 2141+175. Below we mention any evidence that these bona-fide LPRQs in our RoboPol sample had appeared in HPQ state in the old literature, thus indicating a state transition.

{\underline{0014+813}}, {\underline{0833+585}}, {\underline{1606+105}} and 2141+175: These four RoboPol LPRQs were reported to be in the same ({i.e.,} LPRQ) polarization state in the old literature, based on single available measurement. 

{\underline{0333+321}}, 1637+574 and 1928+738: These three RoboPol LPRQs were consistently found in the same ({i.e.,} LPRQ) polarization state in the old literature, based on multiple available measurements. 

0403-132 and 1652+398: These two RoboPol LPRQs are clear cases of state transition from HPQ to LPRQ.

{\underline{0836+710}} and 1426+428: These two RoboPol LPRQs were reported to be in the same ({i.e.,} LPRQ) polarization state in the old literature, based on both available measurements. 

%{\bf (Note: This source has been designated as (LPRQ) as per $p_{new} (max)$). If we take into account the definition using in this section, then this is a LPRQ. Please make me clear regarding the status of this source.}

%1954+513: This confirmed RoboPol LPRQ was reported to be in the same polarization state in the old literature, based on single available measurement.

Thus, it is seen that out of the 11 confirmed LPRQs in the RoboPol sample (Table \ref{tab1}), just 2 are clear cases of polarization state transition over the 3 decades preceding the RoboPol survey. {If we now exclude the 5 of these 11 LPRQs because of their lying at $z > 1$ (whose names are underlined), then the remaining 6 LPRQs contain no clear case of polarimetric state transition.}%, but one probable case of this kind. 

In summary, for our combined RoboPol sample consisting of 59 HPQs, 18 LPRQs and 6 border-line cases {(section \ref{The quasar sample})}, it is seen that only 4 out of these total 83 quasars ($\sim$ 5\%) are clear cases of polarization state transition over the 3 decades preceding the RoboPol survey. The fraction would become 8/83 ($\sim$ 10\%) if the 4 probable cases of transition are also included (section \ref{The RoboPol HPQs}). {Considering only the $z < 1$ subset of our RoboPol sample, which consists of 65 sources, there is only one clear and three probable cases of polarimetric state transition. Thus, the estimates for both the entire RoboPol sample and its $z < 1$ subset are consistent} with the fraction of $\lesssim$ 10\% {reported} in Paper I, based on opto-polarimetric measurements for a 2.6 times smaller sample of RoboPol quasars.

\subsection{Polarization state fluctuations/swings on month/year-like time scales}
Having argued for a remarkable stability of polarization state in an overwhelming majority of radio quasars on decade-like time scales, it would be interesting to verify the previous claims of roughly a quarter of quasars undergoing polarization state transitions on year-like time scales itself, as reported in IT90 (see, also, Paper I). To make an equivalent check employing an independent and larger sample of radio quasars, we use here the polarimetric data listed in Table \ref{tab1} for the present RoboPol sample of {83} radio quasars, where the number of polarimetric measurements {per quasar} has been restricted to 3 (for cases with {available} N > 3, see section \ref{The quasar sample}). Not only is the present sample larger than that used in IT90, it also has the advantage of optical polarimetry performed in a homogeneous manner using a single instrument (RoboPol). From col. 9 of Table \ref{tab1}, out of the total {83} radio quasars, 2 or 3 RoboPol measurements ($p1$, $p2$ and, possibly $p3$) are available for 64 quasars (including 52 HPQs, 10 LPRQs and 2 border-line cases). For these sources, Table \ref{tab1} lists  the highest value of $p_{opt}$, $p1$ followed by $p2$ and $p3$, in decreasing order.
The on-line Fig. S1 displays a plot of $p1$ against $p2$ (or $p3$, if available) for the 64 quasars. {The time interval between the two measurements for each quasar (col. 9 of Table \ref{tab1}) is found to have a median value of $\sim$ 1 year for the sample of 64 quasars.} %A histogram of the time interval  between the two measurements for each quasar (col. 9 of Table \ref{tab1}) is shown in the on-line Figure S3, which yields a median time interval of $\sim$ 1 year for the sample of 64 quasars. 
%{\bf (Note: Out of 64 sources, 52 are HPQs, 10 are LPRQs and 2 are border-line sources)}  
From the on-line Fig. S1, it is seen that {18} out of the {64} quasars ($\sim$ {28} \%) have shown a change of polarization state on the year-like time scales during the RoboPol survey {(2013-17)}, by crossing the $p_{opt} = 3\%$ threshold (in either direction, see Table \ref{tab1}). This fraction is probably an underestimate, if one considers the {8} border-line cases (see on-line Fig. S1) which are consistent with a state transition within 1$\sigma$ error, suggesting that some of them may actually be bona-fide cases of transitions.

Thus, it appears from the analysis presented above that fluctuations/swings of {HP/}blazar state on month/year-like time scales and its long-term persistence (on decade-like time scales) can {occur} in tandem in beamed radio quasars. The two, apparently contrasting behaviours are probably manifestations of different physical processes. In particular, the observed fluctuations of polarization state on month/year-like time scales are probably associated with transient events, like ejections of blobs of nonthermal plasma from the active nucleus (which subsequently appear as VLBI radio components), which too seem to take place on year-like time scales (e.g.,  \citealp{Savolainen2002A&A...394..851S}; \citealp{Lister2009AJ....138.1874L}; \citealp{Liodakis2018MNRAS.480.5517L}). {Clues favouring this scenario emerge from the phenomenology associated with the occurrence of $\gamma$-ray outbursts in FSRQs. These are often found to correlate with (i) millimetric outbursts (e.g., \citealp{Jorstad2001ApJ...556..738J}; \citealp{Lahteenmaki2003ApJ...590...95L}; \citealp{Wehrle2012ApJ...758...72W}; \citealp{Marscher2012arXiv1204.6707M}; \citealp{Alok2017MNRAS.472..788G}) and concomitant variation in optical polarization properties (e.g., \citealp{Blinov2018MNRAS.474.1296B}), and (ii) emergence of new jet features during the outbursts, which subsequently become optically thin and detected by VLBI as knots in the parsec-scale jets (e.g., \citealp{Jorstad2001ApJ...556..738J}). Examples of new plasmon being ejected from the radio core (or passing through it), accompanied by a millimetric/radio (and/or optical) outburst have been extensively reported  
(e.g., \citealp{Krichbaum1990A&A...237....3K}; \citealp{Mutel1990ApJ...352...81M}; \citealp{Abraham1996A&AS..115..543A}; \citealp{Tateyama1999ApJ...520..627T}; \citealp{Savolainen2002A&A...394..851S}; \citealp{Karamanavis2016A&A...586A..60K}; \citealp{Liodakis2020ApJ...902...61L}).} 
%Examples of new plasmon ejections from the core, coinciding with a millimetric/radio/optical outburst have been noted for a long time (e.g.,  \citealp{Krichbaum1990A&A...237....3K}; \citealp{Mutel1990ApJ...352...81M}; \citealp{Abraham1996A&AS..115..543A}; \citealp{Tateyama1999ApJ...520..627T}; \citealp{Savolainen2002A&A...394..851S}; \citealp{Karamanavis2016A&A...586A..60K}).}
%Examples of such plasmon injection coinciding with millimetric/radio flare have been noted for a long time (e.g., \citealp{Savolainen2002A&A...394..851S}; also, \citealp{Abraham1996A&AS..115..543A}; \citealp{Krichbaum1990A&A...237....3K}; \citealp{Mutel1990ApJ...352...81M}; \citealp{Tateyama1999ApJ...520..627T}; \citealp{Karamanavis2016A&A...586A..60K}).} 

\section{Conclusions}
\label{conclusions}
For a well-defined sample of {83} radio quasars extracted from the RoboPol survey (2013-17), for which opto-polarimetric measurements {taken prior to 1990} could be found in the literature, we have made a source-wise comparison of the polarization states observed during these two time spans separated by about 3  decades, taking the conventional $p_{opt}$ = 3\% as the { division} between HPQs (blazars) and LPRQs. Our {well-defined} sample consists of {59} HPQs, {18} LPRQs and {6} border-line cases. This comparison, based purely on opto-polarimetric measurements for a fairly large quasar sample, has revealed that $\sim$ 90\% of them have retained their polarization class over the 3-4 decades long time baseline. This confirms the finding of Paper I, based on opto-polarimetric data for a $\sim$ {2.5} times smaller RoboPol sample of quasars. {The implied long-term stability of optical polarization state in radio quasars is shown here to hold even for their $z < 1$ subset whose optical flux is much less contaminated by the thermal emission from the accretion disk.} The long-term stability is particularly remarkable in view of the polarization state 
swings/{transitions} which are also found here and elsewhere (IT90; Paper I) to occur on year-like time scales, for a sizable fraction ($\sim$ 25\%) of radio quasars. Apparently, the long-term stability of the {HP/blazar} mode and its transience on much shorter (year-like) time scales comprise two physical processes which need to be distinguished, although they might be at work simultaneously. The fluctuations/swings of polarization state on year-like time scales are probably associated with transient events, such as the relatively frequent, semi-regular ejections of blobs of synchrotron plasma {from the central engine}, which are subsequently observed as parsec-scale radio knots in the VLBI images of the nuclear radio jets.

\section*{Acknowledgements}
{We thank the anonymous reviewer for a constructive feedback.}
GK acknowledges a Senior Scientist fellowship from the Indian National Science Academy. This research has made use of data from the RoboPol programme, a collaboration between Caltech, the University of Crete, IA-FORTH, IUCAA, the MPIfR, and the Nicolaus Copernicus University, which was conducted at Skinakas Observatory in Crete, Greece.
This work has also made use of the NASA/IPAC Extragalactic Database (NED) which is operated by the Jet Propulsion Laboratory, California Institute of Technology, under contract with the National Aeronautics and Space Administration.

\section*{Data availability}
The data used in this study are publicly available in peer-reviewed publications listed in the reference section.

%\clearpage

%%%%===========================================================

%\clearpage
%\bibliographystyle{aa} 
\bibliography{references}

\end{document}